\newif\ifAnon\Anonfalse

\ifAnon 
\documentclass[journal=tches,submission,hyperref=pagebackref=true]{iacrtrans}
\else
\documentclass[journal=tches,preprint,hyperref=pagebackref=true]{iacrtrans}
\fi

\usepackage{csquotes}
\usepackage{booktabs}
\usepackage{wasysym}
\usepackage{listings}

\usepackage{caption}
\usepackage{subcaption}
\usepackage{pifont}

\author{
Luca Wilke \and
Jan Wichelmann \and
Anja Rabich \and
Thomas Eisenbarth
}
\institute{
  University of Lübeck, Lübeck, Germany
        \email[l.wilke@uni-luebeck.de,j.wichelmann@uni-luebeck.de,a.rabich@uni-luebeck.de,thomas.eisenbarth@uni-luebeck.de]{{l.wilke,j.wichelmann,a.rabich,thomas.eisenbarth}@uni-luebeck.de}

}

\title[SEV-Step: A Single-Stepping Framework for AMD-SEV]{SEV-Step\\ A Single-Stepping Framework for AMD-SEV}

\begin{document}

\maketitle


\begin{abstract}
The ever increasing popularity and availability of Trusted Execution Environments (TEEs) had a stark influence on microarchitectural attack research in academia, as their strong attacker model both boosts existing attack vectors and introduces several new ones.
While many works have focused on Intel SGX, other TEEs like AMD SEV have recently also started to receive more attention.
A common technique when attacking SGX enclaves is single-stepping, where the system's APIC timer is used to interrupt the enclave after every instruction. Single-stepping increases the temporal resolution of subsequent microarchitectural attacks to a maximum.
A key driver in the proliferation of this complex attack technique was the SGX-Step framework, which offered a stable reference implementation for single-stepping and a relatively easy setup.
In this paper, we demonstrate that SEV VMs can also be reliably single-stepped. To lay the foundation for further microarchitectural attack research against SEV, we introduce the reusable SEV-Step framework. Besides reliable single-stepping, SEV-Step provides easy access to common attack primitives like page fault tracking and cache attacks against SEV. All features can be used interactively from user space.
We demonstrate SEV-Step's capabilities by carrying out an end-to-end cache attack against SEV that leaks the volume key of a LUKS2-encrypted disk. Finally, we show for the first time that SEV is vulnerable to Nemesis-style attacks, which allow to extract information about the type and operands of single-stepped instructions from SEV-protected VMs.

\end{abstract}

\section{Introduction}
Microarchitectural side-channel security of computer systems has been one major 
pillar of computer security research in recent years.
In microarchitectural attacks, the adversary aims to infer/extract secret information through observations of the system's microarchitectural state.
With the ever increasing popularity and availability of Trusted Execution Environments (TEEs), side-channel attacks are more relevant than ever, as the attacker model of TEEs includes powerful system-level attackers. Naturally, such an attacker has more capabilities to observe the system's microarchitectural state, extending the potential attack surface. While attacks targeting TEEs build on a variety of data sources, like cache state, microarchitectural buffers or power reporting interfaces, they share the property that they have to synchronize their data sampling with the execution flow of the victim. For example, monitoring the cache state only leaks meaningful information if the victim is about to perform a vulnerable memory access.
An increased temporal or spatial resolution of the attacker's ability to infer the victim's execution state often drastically improves the amount/quality of leaked data.

One commonly used technique with both Intel SGX and AMD SEV is disabling certain memory pages, such that the victim is forced to handle a page fault when it tries to access those pages. This allows the attacker to synchronize with the victim's execution flow on a page-granular level.
For Intel SGX, researchers tried to further increase the resolution by interrupting the SGX enclave with a high frequency, e.g., by using the system's APIC timer. Eventually, Van Bulck et al.\ demonstrated in SGX-Step~\cite{DBLP:conf/sosp/BulckPS17} that an attacker can even achieve the maximum temporal resolution of interrupting the victim after every single instruction. However, besides this technical improvement over prior work that was only able to interrupt SGX enclaves every few instructions, they were also the first to introduce a reusable framework. Now, at the time of writing this paper, SGX-Step has been used in 33 publications~\cite{sgxstepgithub}, clearly showing the benefits of reusable building blocks in a research area where the technical challenges and nuances are ever increasing. 

While, for example, page fault tracking is also commonly used in SEV, most prior work has either not released any artifacts at all \cite{DBLP:conf/uss/LiZWLC21,DBLP:conf/sp/MorbitzerPRDS21,DBLP:conf/ccs/LiZL21,DBLP:conf/uss/LiZLS19,DBLP:conf/ccs/WernerMAPM19,DBLP:journals/corr/abs-1712-05090} or artifacts that are highly specific to the demonstrated attack \cite{DBLP:conf/sp/WilkeWM020,DBLP:conf/sp/WilkeWS021}.
Exceptions to this are~\cite{DBLP:conf/eurosys/Morbitzer0HW18} and~\cite{DBLP:conf/sp/LiWW0TZ22}.
The framework from~\cite{DBLP:conf/eurosys/Morbitzer0HW18} allows to track pages accessed by the SEV VM as well as remapping pages, but only applies to the first two versions of SEV. In addition, it does not allow to interactively react to page faults in a synchronous manner,  making it unsuitable for many types of side-channel attacks.
While the framework from~\cite{DBLP:conf/sp/LiWW0TZ22} offers such interactivity, it also is restricted to page fault granularity.

\medskip
\noindent
\textbf{Our Contribution} 
is twofold: First, we introduce \emph{reliable single-stepping in the context of SEV}\footnote{
    Concurrent to our work, PwrLeak~\cite{DBLP:conf/dimva/WangLZL23} also uses single-stepping against SEV VMs, but neither discusses reliability nor do they introduce a reusable framework.
}.
The second contribution is making interactive single-stepping, page fault tracking and eviction set-based cache attacks available in a \emph{single, reusable framework}. 
Our framework shifts most of the complex attack logic from kernel space into user space, allowing the development of new attacks entirely with user space code.
In the hope that the framework inspires a similar community as SGX-Step, we dubbed it SEV-Step.

Furthermore, to showcase the capabilities of our framework as well as its academic relevance, we demonstrate an end-to-end key extraction attack against a SEV VM and utilize SEV-Step to detect and quantify instructions based on their execution time. 
The end-to-end cache attack succeeds in extracting a LUKS2 disk encryption key from a SEV-protected VM using a single trace.
The SEV-Step-based instruction latency analysis confirms that an attacker can leak information about the type and operands of certain instructions in SEV by measuring the time required for single-stepping them. Such classification was previously shown for SGX by Van Bulck et al.\ in Nemesis~\cite{DBLP:conf/ccs/BulckPS18}.

\noindent In summary, this work
\begin{itemize}
    \item introduces reliable single-stepping against SEV VMs;
    \item provides a reusable framework facilitating future attack research against SEV;
    \item steals disk encryption keys in an end-to-end cache attack; and
    \item shows SEV's vulnerability to Nemesis-style~\cite{DBLP:conf/ccs/BulckPS18} attacks.
\end{itemize}

\noindent The remainder of the paper is organized as follows:
\autoref{sec:background} provides background about relevant x86 system architecture and SEV. \autoref{sec:sev-step-design}
starts with a general overview over the SEV-Step framework, before explaining its implementation in detail. Next, \autoref{sec:evaluation} evaluates the single-stepping and cache attack features of the framework.
Finally, \autoref{sec:case-studies} demonstrates an end-to-end cache attack stealing disk encryption keys and shows that SEV is vulnerable to Nemesis-style~\cite{CCS:BulPieStr18} attacks.

\section{Background}
\label{sec:background}
\subsection{AMD SVM}
\label{sec:background-svm}
\begin{figure}
    \centering
    \includegraphics[width=0.45\textwidth]{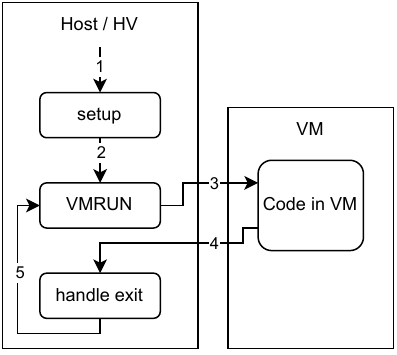}
    \caption{Basic control flow of a hypervisor using hardware assisted virtualization on AMD. After some initial setup (1), the hypervisor enters the main control loop. The \texttt{VMRUN} instruction takes care of performing the context switch into the VM (3). Afterwards, the VM is running until a \texttt{VMEXIT} event occurs (4), upon which the hardware restores
    the host context and resume the execution immediately after the \texttt{VMRUN} instruction (4) where the exit is handled, before the VM is entered again.}
    \label{fig:basic-hv}
\end{figure}

AMD Secure Virtual Machines (SVM) is AMD's instruction set extension for hardware-accelerated virtualization.
It introduces the concepts of \textit{guest mode} and \textit{host mode}. Both modes have the full set of privilege levels of the x86 architecture. However, in guest mode certain instructions have slightly different semantics in order to enable the virtualization concept. As shown in \autoref{fig:basic-hv}, from host mode, we can enter the guest mode using the \texttt{VMRUN} instruction. The
CPU runs in guest mode until an intercepted event occurs, which leads to a \texttt{VMEXIT}, returning the execution flow to the instruction immediately following the \texttt{VMRUN} instruction used to enter the VM. For both, VMRUN and VMEXIT, the
hardware takes care of storing/restoring the current context, like the register values.

The host mode can pass a configuration struct called VMCB to the VMRUN instruction to configure, among other things, which events lead to a VMEXIT. This interception mechanism enables the host/hypervisor (HV) to transparently simulate certain
behavior to the guest. Furthermore, the mechanism enables the HV to ensure that it stays in control of the hardware by causing periodic VMEXITs through APIC timer interrupts. Interrupt handling is discussed in detail in the next section.

SVM also introduces the concept of nested page tables (NPT) easing the virtualization of memory. With NPT, VM can no longer address real physical memory with its page tables. Instead the memory subsystem uses a second set of page tables, the hypervisor-controlled NPT, to translate the so-called Guest Physical Addresses (GPA) of the VM's page table to real physical addresses.

\subsection{APIC and APIC Timer}
\label{sec:background-apic}
According to the AMD Programmer's Manual~\cite[Sec. 16]{amdProgrammersManualVol2}, the Advanced Programmable Interrupt Controller
(APIC) is located between the CPU core and the rest of the system. It is responsible for
providing the CPU core(s) with interrupts. Those interrupts can either originate from sources local to the APIC, like the APIC
timer interrupt, or from sources remote to the APIC, e.g., from the IOAPIC.
\label{sec:apic-timer-background}

The APIC timer is part of the APIC Controller. It is a counter that is decremented with a configurable frequency. Once it reaches zero, it generates an interrupt. It can either be used in oneshot mode or in periodic mode. The latter restarts the timer once it reaches zero, while the former does not. The APIC timer is commonly used by the OS to implement periodic jobs like process scheduling~\cite[Sec. 16.4.1]{amdProgrammersManualVol2}.

\subsection{Interrupt Handling in AMD SVM}
\label{sec:background-interrupts-svm}
Under the x86 architecture, the delivery of interrupts is controlled via the \texttt{EFLAGS.IF} field.
If set to 0, interrupt delivery is suppressed. This is called \emph{masking} an interrupt. Masked interrupts are held waiting/pending until \texttt{EFLAGS.IF} is set to 1 again\cite[Sec. 8.1.4]{amdProgrammersManualVol2}.

In contrast to exceptions or traps, interrupts are inherently asynchronous to the currently executing program. However, instead of immediately aborting program execution, they are only processed on \textit{instruction boundaries}, meaning that the currently executing instruction will still be retired before the interrupt is handled~\cite[Sec 8.2.24]{amdProgrammersManualVol2}.

When using AMD's SVM to run a virtual machine, we distinguish between physical interrupts and virtual interrupts.
Physical interrupts are interrupts that are actually generated by the hardware. As discussed in the previous section,
the HV can configure the VMCB such that certain interrupts lead to a VMEXIT, returning control from the guest mode to the HV.
However, to facilitate virtualization, the HV may decide to \enquote{pass on} the interrupt to the VM as a \textit{virtual interrupt}. This mechanism is called \emph{interrupt injection} and is performed via configuration fields in the VMCB.

To ensure that the VM cannot simply mask all physical interrupts using its version of the EFLAGS.IF register, 
the HV can configure the VMCB such that
the VM's EFLAGS.IF flag only affects virtual interrupts. This way, the VM cannot prevent actual physical interrupts from being delivered\cite[Sec. 15.21]{amdProgrammersManualVol2}.

\subsection{AMD SEV}
AMD Secure Encrypted Virtualization (SEV)~\cite{plainSevWhitepaper} is a Trusted Execution Environment (TEE) protecting whole virtual machines from a malicious HV and to some extent against physical attackers.
It builds on the AMD SVM hardware acceleration for virtualization. These kinds of TEEs are also known as confidential VMs.
With SEV, each VM's memory content is encrypted with AES-128 using the XOR-Encrypt-XOR (XEX)~\cite{AC:Rogaway04} mode before leaving the main processor. A dedicated co-processor, the AMD Platform Security Processor (PSP), forms the root of trust of the system. It takes care of securely handling the memory encryption keys and offers an API to the HV to setup and manage SEV VMs.
While located inside the main processor, for example in the cache, each VM's data is assigned a different tag, called Address Space Identifier (ASID) to ensure isolation.
After the initial release of SEV, there were two iterative enhancements called SEV-ES~\cite{kaplan:2017:seves} and SEV-SNP~\cite{sevSNPWhitepaper}, the latter being the latest version.

\subsection{Attacks on AMD SEV}
Since its release, there has been a long line of attacks against AMD SEV.

\noindent\textbf{Unencrypted VMCB}: In
\cite{DBLP:conf/vee/HetzeltB17,DBLP:conf/ccs/WernerMAPM19} the authors exploit the unencrypted VM register state inside the VMCB, which has been mitigated with SEV-ES.

\noindent\textbf{Nested Page Tables}: In \cite{DBLP:conf/vee/HetzeltB17,DBLP:conf/eurosys/Morbitzer0HW18,DBLP:conf/sp/MorbitzerPRDS21,DBLP:conf/codaspy/Morbitzer0H19} the authors exploit the HV's control over the nested page tables to remap pages either leaking data or injecting code. These attacks are mitigated with SEV-SNP.

\noindent\textbf{Encryption Mode}:
In~\cite{DBLP:conf/codaspy/BuhrenGNSV17} the attacker exploits the unauthenticated encryption to fault computations inside the VM by flipping ciphertext bits. \cite{DBLP:journals/corr/abs-1712-05090,DBLP:conf/sp/WilkeWM020} reverse engineer the encryption mode together with the tweak values and show how this can be used to leak or inject data into the VM. However, on more recent EPYC CPUs, the updated XOR-Encrypt-XOR (XEX)~\cite{AC:Rogaway04} mode prevents the tweak reverse engineering, and SEV-SNP additionally prohibits writes to the VM's memory.
\cite{DBLP:conf/uss/LiZLS19} show that the bounce buffers required for I/O interaction between HV and VM in combination with the insufficient binding of ciphertext to its memory location prior to the XEX mode can be used to leak/inject data.
Finally, \cite{DBLP:conf/uss/LiZWLC21,DBLP:conf/sp/LiWW0TZ22} demonstrate than even with SEV-SNP, the attacker can still exploit the fact that the memory encryption mode is deterministic to leak data through a side-channel.

\noindent\textbf{Miscellaneous}:
In~\cite{DBLP:conf/dimva/WangLZL23} the authors exploit the software-accessible power reporting features on AMD CPUs to unveil the type of executed instructions. However, the attack was only demonstrated with plain SEV, and the applicability to more recent versions is uncertain.
In~\cite{DBLP:conf/ccs/BuhrenWS19,DBLP:conf/ccs/BuhrenJKS21} Buhren et al. show hardware-based power glitching attacks against SEV's root of trust, the Platform Security Processor (PSP), granting them custom code execution on the PSP.
Further attacks on SEV versions prior to SEV-SNP also exploited flaws in the ASID-based isolation~\cite{DBLP:conf/ccs/LiZL21}, in the calculation of the attestation value~\cite{DBLP:conf/sp/WilkeWS021}, as well as in the software interface between HV and VM~\cite{DBLP:journals/corr/abs-2010-07094}.

\subsection{Interrupt-Based Single Stepping}
The idea of of improving the temporal resolution of microarchitectural
attacks via triggering frequent interrupts was first explored in the context
of SGX. There are several works~\cite{DBLP:conf/ches/MoghimiIE17,DBLP:conf/usenix/HahnelCP17,DBLP:conf/uss/0001SGKKP17} that significantly improved the temporal resolution from the page fault level down to a few instructions. However, reliable single-stepping was only achieved with SGX-Step~\cite{DBLP:conf/sosp/BulckPS17}.
The idea of APIC timer-based stepping was first applied to SEV in Cipherleaks~\cite{DBLP:conf/uss/LiZWLC21}. However, they \emph{did not achieve reliable single-stepping} (c.f. Figure 3 in~\cite{DBLP:conf/uss/LiZWLC21}) and did not publish any code artifacts.

Concurrent to our work, PwrLeak~\cite{DBLP:conf/dimva/WangLZL23} 
also uses APIC timer-based single-stepping. However, they only performed their experiments on the outdated, plain SEV variant (not on SEV-ES or SEV-SNP) and do not discuss reliability. They committed to open-sourcing their attack code. At the time of writing, though the final version of the PwrLeak paper has been published, the code is not available yet.

\subsection{Cache Attacks}
\label{sec:cache-attacks-background}

Since CPUs are much faster than main memory, they use caches to store recently accessed data in order to minimize latency. Modern CPUs usually use set-associative caches, where each memory address maps to a specific location in the cache, called the \emph{cache set}. Each cache set has a limited amount of slots to store data, called \emph{ways}. If all ways of a cache set are used, new data will evict one of the older entries.
Each cache entry is identified via a unique tag value.
Cache attacks use timing to infer whether a certain address is currently cached or not. As shown in many works~\cite{percival2005cache,bernstein2005cache,DBLP:conf/sp/LiuYGHL15,DBLP:conf/sp/ApececheaES15}, an attacker can use this to leak secrets from other processes/entities on the system.

In \emph{Prime+Probe}~\cite{DBLP:conf/ctrsa/OsvikST06,DBLP:conf/sp/LiuYGHL15}, the attacker accesses a specifically crafted set of memory addresses, a so-called \emph{eviction set}, to fill up a cache set.
Next, the attacker waits for the victim to perform a memory access.
Finally, the attacker accesses the eviction set again, measuring the required time. A long access time indicates that the victim's memory access mapped to the same set, and thus evicted one of the attacker's entries.

The \emph{Load+Reload}~\cite{ASIACCS:LHSPMG20} attack is a more recent variation of the Prime+Probe attack. It exploits a specific behavior of the way predictor present on AMD CPUs since the Bulldozer microarchitecture: Accesses to the same
physical address but with different virtual addresses always encounter a L1 data cache miss. This allows an attacker to perform the Prime+Probe step using only a single memory access for each step, irrespective of the number of ways the cache has.

Another popular cache attack is the \emph{Flush+Reload}~\cite{DBLP:conf/uss/YaromF14} attack.
Like with Load+Reload, Flush+Reload 
requires shared memory between the attacker and the victim.
First, the attacker uses an architectural flush command, like \texttt{clflush}, to remove the shared data/code from the cache. As with the other techniques, the attacker waits for the victim to execute. To probe if the victim has accessed the memory location, the attacker finally measures the time required to access the flushed data with his mapping.

\section{Attacker Model}
In this paper, we assume a software-level attacker with full system-level privileges, which matches the threat model of AMD SEV. Using these capabilities, the attacker acts as a malicious hypervisor running a modified Linux kernel. Furthermore, the attacker can freely tweak nearly all system settings, like fixing the CPU frequency or disabling hardware cache prefetchers. However, a few features, like
the availability of simultaneous multi threading (SMT) or the firmware version of the root of trust are part of the attestation report~\cite{amdSnpAbi}. Thus their configuration status is visible to the VM owner.
The attacked VMs are protected with AMD SEV-SNP. Due to SEV's attestation feature, the software inside the VM is assumed to be benign and under the VM owner's control.

\section{SEV-Step Design}
\label{sec:sev-step-design}
In this section, we first motivate the design of SEV-Step and its components, and then describe each component in-depth.
The framework consists of the following main components: Single-stepping,
page fault tracking and eviction set-based cache attacks.

\subsection{Design Goals}
We identified two major design goals for SEV-Step: Interactivity and reusability.

\bigskip\noindent\textbf{Interactivity:} One key component for side-channel attacks in general is to precisely link the (micro)architectural observations with the victim's execution state.
In the context of TEEs, like Intel SGX or AMD SEV, this is commonly achieved by interrupting the victim at defined points in its execution state, allowing the attacker to either prepare or sample the (micro)architectural state.
Thus, the SEV-Step framework should not only allow the attacker to interrupt the VM, but also notify the attacker about the interruption, keeping the VM paused until the attacker signals that they are ready for the VM to resume.

\bigskip\noindent\textbf{Reusability:} Since features like page fault tracking or programming the APIC timer require the use of certain privileged OS resources, it is natural to implement them directly inside the OS kernel. However, patching the kernel comes with several downsides. Small errors can easily lead to system crashes or hard-to-debug instabilities. Furthermore, the programming environment is limited to C, without any external libraries.
Finally, recompiling the Linux kernel is quite resource-intensive, leading to long iteration times during development.
Thus, we aim for a design that only implements the basic primitives that are dependent on privileged OS resources inside the kernel. These primitives are then made available to a user space library via an API allowing the development of complex attack logic in the richer and less error-prone programming environment available to user space code. 
Given our first goal of interactivity, this requires us to build a synchronous, bidirectional channel between the kernel space and the user space components.
In addition, bundling the API in a separate library also makes it easy to separate attack specific logic from the framework code itself. This is showcased in the end-to-end attack in \autoref{sec:aes-attack}, which is a completely separate code base that only links to the SEV-Step library.

\subsection{User Space API}
\begin{figure}[t]
    \centering
    \includegraphics[width=0.5\textwidth]{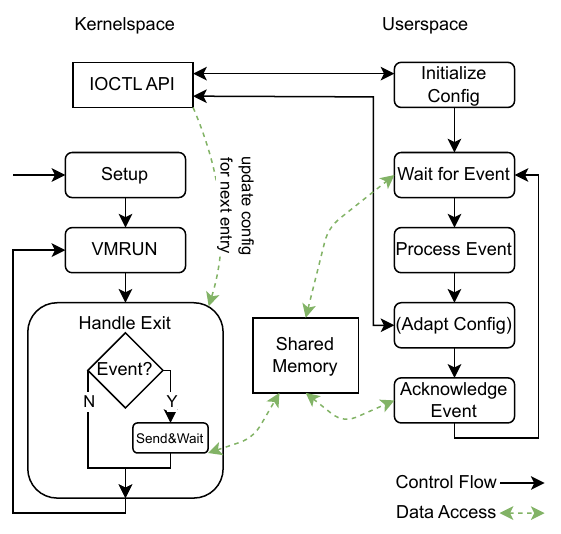}
    \caption{Overview of the kernel space and user space parts of the SEV-Step framework. There are two communication channels: An ioctl API, and communication over shared memory.
    Sending and acknowledging (single-stepping) events is done over shared memory. Upon sending an event, the kernel space part blocks until the event is acknowledged, delaying the next VMRUN. Both waiting for new events and for acknowledgments are implement via active polling to reduce latency.
    As changes to the VM can only be made upon the next exit, the ioctl API only updates
    a central configuration struct, deferring the application of the changes to the next exit. However, in combination with the blocking event handling, the user space library can
    synchronize these changes to the VM state.
    }
    \label{fig:sev-step-workflow}
\end{figure}

We built SEV-Step on top of AMD's reference hypervisor implementation for SEV, which is based on the Linux KVM kernel module and QEMU.

\autoref{fig:sev-step-workflow} shows an overview of the interaction between user space and kernel space in SEV-Step, as well as the basic workflow of the framework.
The left-hand side shows the kernel space part, while the right-hand side shows the user space part.
There are two communication channels between the kernel space and the user space part: ioctls and shared memory.

Ioctls are a commonly used approach to implement kernel space to user space APIs. An ioctl is a basically a wrapper system call, that can be filled with custom behavior.
However, being a system call, they require a full user space to kernel space context switch.
In addition ioctls do not allow the kernel space to push events to user space.
Thus, we only use ioctls for low-frequency operations, like initialization or configuration.
For the high-frequency page fault and single-step event notifications, we use a custom, lightweight protocol over shared memory.

As explained in \autoref{sec:background-svm}, the core part of the KVM hypervisor kernel module is a control loop
around the VMRUN instruction. For the SEV-Step framework, we mainly add additional control logic before
and after the VMRUN instruction, that, e.g., primes/probes the cache or starts the APIC timer. In addition, we also need to patch KVM's page fault handling code and overwrite the default APIC timer handling.
This additional control logic can be configured via the ioctl API.
As we can only reconfigure the VM between VMRUNs, the ioctl API inherently is not synchronized with the control loop, i.e., changes only take effect
on the next entry/exit from the VM.
While this seems to contradict the interactivity design goal, the situation can be resolved by the blocking event notification mechanism explained in the next paragraph.

When a VMEXIT occurs due to a single-step or page fault event, the kernel space
part uses the shared memory channel to deliver an event to the user space counterpart.
However, after sending the event, the kernel space does not continue the execution of the VM, but instead waits for the user space to acknowledge the event, keeping the VM in a paused state. This enables the user space part to make configuration changes via the ioctl API that immediately take effect on the next VMRUN. In addition, the semantics conveyed by the page fault/single-step events allow the user space application to deduce the internal state of the VM, as required by the interactivity design goal.

To synchronize the memory accesses to the shared memory area, both sides actively poll a spin-lock. Compared to, e.g., mutexes, which might lead to an immediate reschedule when encountering an already taken lock, this results in lower overhead.

\subsection{Single Stepping}
This section describes how single-stepping is implemented in the SEV-Step framework.
We start by describing the basic mechanism before giving more details on tweaking the mechanism to achieve reliable single-stepping.

First, the HV uses the VCMB configuration structure, which is passed to the VMRUN instruction when entering the VM, to ensure that an APIC timer interrupt leads to a VMEXIT (c.f. \autoref{sec:background-interrupts-svm}). Next, the HV programs the APIC timer and enters the VM with the VMRUN instruction. Ence the timer expires, the resulting interrupt results in a VMEXIT, handing control back to the HV.
This workflow is part of the HV's regular operations, as it uses the APIC timer anyway to implement a periodic tick/callback.
Next, we discuss how to use this mechanism to achieve single-stepping.

\begin{figure}[t]
    \centering
    \includegraphics[width=0.5\textwidth]{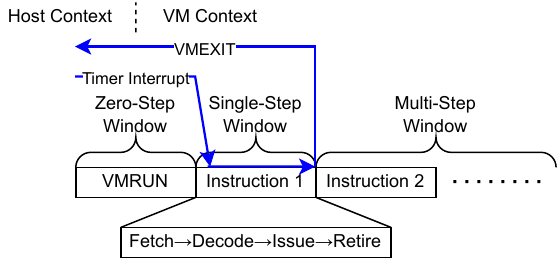}
    \caption{Timeline of the executed instructions during a HV to VM context switch. As the APIC timer interrupt is only processed at instruction boundaries, we get timing windows instead of discrete points in time, at which the interrupt leads to zero-, single- or multi-steps. The bottom row depicts that, internally, the execution of an instruction consists of several stages.}
    \label{fig:stepping-intervals}
\end{figure}

As explained in~\autoref{sec:background-interrupts-svm}, the hardware does not immediately trigger a VMEXIT upon receiving, e.g., a timer interrupt. Instead, the interrupt handling, and thus the VMEXIT, is postponed until the next instruction boundary is reached.
As shown in \autoref{fig:stepping-intervals}, to achieve single-stepping, we need to configure the timer such that the interrupt is triggered before the first instruction in the VM's execution flow is finished. However, the timer interval also needs to be long enough for the first instruction of the VM to be issued into the execution pipeline. Otherwise, the VM would exit without having executed a single instruction. If the APIC timer interval is too large, multiple instructions are executed. We call these events, single-, zero- and multi-step, respectively.

\subsubsection*{Increasing Single Step Time Window}
\label{sec:step-window}
As shown in the bottom half of \autoref{fig:stepping-intervals}, the execution of an instruction can be decomposed into multiple parts. While the time between issuing and retiring an instruction
might be very short, e.g., when executing a \texttt{nop} instruction, the other steps still require some time. This is the case especially when the CPU needs to fetch the instruction from memory and, if applicable, resolve other memory addresses used by the instruction.
In our experiments, we found that the size of the single-step window is not dominated by the instruction's type itself, but rather by 
the instruction-agnostic execution stages (fetch, decode, $\dots$). To enable reliable single-stepping, an attacker must ensure that the execution of the VMRUN instruction always takes roughly the same amount of time and that the single-step window never drops below a certain threshold.

In order to maximize the time required for the first instruction executed inside the VM, we explored flushing 
the VM's Translation Lookaside Buffer (TLB) entries as well as resetting the \enquote{accessed} bit~\cite[Sec. 5.4.1]{amdProgrammersManualVol2} of the page containing the first instruction that would be executed after the VMRUN.
By flushing the VM's TLB entries, we ensure that accessing the code page that contains the first instruction always requires a time-consuming page table walk to translate the address. The same applies to all memory operands used by the instruction.
The intention behind resetting the \enquote{accessed} bit is similar.
If cleared, the hardware has to set the accessed bit again~\cite[sec. 5.4.2]{amdProgrammersManualVol2}.
According to the Intel SGX-specific AEX-Notify~\cite{aexNotify} paper, this requires substantial time and is one of the key factors for reliable single-stepping on Intel SGX.
We evaluate the effects in \autoref{sec:eval-step-reliability}.

Finally, we tweak the system configuration as follows to ensure a stable execution speed.
We pin the kernel thread running the VM to a dedicated CPU core, that does not run any other tasks. This is implemented via the \texttt{isolcpus}, \texttt{nohz\_full}, \texttt{rcu\_nocbs} and \texttt{rcs\_nocb\_poll} Linux kernel parameters~\cite{linuxKernParametersManual}.
In addition, we ensure a stable CPU frequency by either disabling dynamic frequency scaling altogether (if the BIOS permits it), or by pinning the CPU frequency using the Linux \texttt{cpufreq} subsystem~\cite{linuxKernCpufreq}.
Finally, we disabled hardware cache prefetchers in the BIOS.
Since SEV aims to protect against a privileged system-level attacker, all of these changes are within the threat model.

\subsubsection*{Preventing Virtual Timer Interrupts}
\label{sec:cancel-injected-interrupts}
As explained before, the Linux OS uses the APIC timer to implement a periodic tick/callback.
Thus, while the HV handles the physical APIC timer interrupts, it also needs to emulate the interrupt for the VM. As explained in \autoref{sec:background-interrupts-svm}, AMD's hardware assisted virtualization offers the concept of virtual interrupts to achieve this. Thus, whenever the APIC timer interrupts the VM, the KVM hypervisor would usually inject a virtual timer interrupt into the VM upon the next VMRUN. As a consequence, the Linux OS in the VM jumps to its corresponding interrupt handler. As a result, an attacker would not single-step any user code, but only the VM's APIC timer interrupt handler. Thus, we need to modify this part of KVM's logic to prevent any virtual timer interrupt injection while we single-step the VM. For our attacks, we did not observe any instabilities in the VM's execution due to the inhibited interrupt. As a workaround for potential issues with very long single-step phases, we could periodically allow the injection of the virtual timer interrupt.

\subsubsection*{Determining the Step Size}
\label{sec:determining-step-size}
To properly determine the APIC timer timeout value, we need a feedback channel enabling us to observe the amount of instructions executed by the guest. In SGX-Step~\cite{DBLP:conf/sosp/BulckPS17} the \enquote{accessed} bit of the page table entry corresponding to the page containing the current instruction is used to differentiate single-steps and zero-steps. However, it cannot be used to detect multi-steps, which is only possible by running the enclave in debug mode to observe its instruction pointer.
While these two methods also work in SEV, we additionally have access to the VM's performance counter events. As demonstrated in~\cite{DBLP:conf/sp/LiWW0TZ22}, there is a performance counter for retired instructions that can be configured to only consider instructions executed by the VM. Thus, evaluating the counter before and after entering the VM immediately reveals the step size.

\subsection{Page Fault Tracking}
For page fault tracking, SEV-Step uses the well-known control of the HV over the nested page tables~\cite{DBLP:conf/eurosys/Morbitzer0HW18,DBLP:conf/ccs/LiZL21,DBLP:conf/sp/WilkeWM020}.
By modifying the \emph{present}, \emph{no-execute} and \emph{read}/\emph{write} bits of a page, the HV can force the VM to encounter a page fault that also reveals the type of access.
While being more coarse-grained than single-stepping, page fault-based tracking is significantly faster. Thus, for many attack scenarios, it is beneficial to rely on the coarse-grained page fault mechanism as much as possible before enabling single-stepping. For example, an attacker could use page fault tracking to get notified when the VM is about to execute a code page containing
a series of secret-dependent memory lookups. Only then, the attacker activates single-stepping,
allowing them to, e.g., perform a cache attack against each individual memory access.

\subsection{Cache Attacks}
\label{sec:cache-attacks}
To use SEV-Step's cache attack capabilities, the attacker first needs to perform some initial configuration like locating and defining the cache attack targets. Afterwards, while single-stepping the VM, the attacker can request that a cache attack is performed for the next single-step. The resulting step event is enriched with the measured data.

In the remainder of this section, we discuss how we measured execution times on our system as well as the applicability of the Prime+Probe, Load+Reload and Flush+Reload (c.f. \autoref{sec:cache-attacks-background}) cache attacks in the context of SEV.

\subsubsection*{Measuring Access Times}
As, e.g., discussed in \cite{ASIACCS:LHSPMG20}, the \texttt{rdtsc} and \texttt{rdtscp} instructions return very coarse-grained timing data on AMD CPUs since the Zen microarchitecture. This makes them unsuitable for cache attacks without averaging over several iterations.
Prior work suggests either using a so-called counting thread~\cite{DBLP:conf/uss/LippGSMM16} or the \texttt{rdpru} instruction~\cite{amdProgrammersManualVol3,DBLP:conf/uss/LippG022,ASIACCS:LHSPMG20}.
While the latter could be disabled for unprivileged users, we assume an attacker with kernel-level privilege. In the following sections, we use the \texttt{rdpru} instruction due to the lower footprint on the microarchitectural state compared to the counting thread.

In addition to measuring the access time, we can also use performance counters to gather information about the cache state. As described earlier (c.f. \autoref{sec:determining-step-size}),
SEV does not offer protection/isolation for performance counters.
For the level 2 (L2) cache, there are performance counters for \enquote{L2 Cache Miss from L1 Data Cache Miss} and \enquote{L2 Cache Hit from L1 Data Cache Miss}~\cite[Sec. 2.1.17.2]{amdProcessorProgReferenceFam19hM01hB1}. 
However, as there is no performance counter for L1 data cache hits or misses, we still require the access time to infer the L1 result in order to interpret the L2 events. E.g., if we have a L1 cache hit, the difference in both counters would be zero, leaving us with an inconclusive result until evaluating the access time.

\subsubsection*{Flush+Reload}
The HV can easily obtain a mapping to any of the VM's memory pages by using the nested page tables.
However, as explained in~\cite{DBLP:conf/ccs/LiZL21}, in SEV the cache tag is extended with the current ASID and the encryption status of the corresponding page (C-Bit), effectively allowing the same data to reside in the cache multiple times. As the HV has a different ASID than the VM (as discussed in ~\cite{DBLP:conf/ccs/LiZL21} the HV could technically change its ASID, but this would basically prevent it from executing any further code), it cannot get a hit on the data brought into the cache by the VM, when accessing the data via its own mapping. Thus, the HV cannot perform the reload part of the Flush+Reload attack.

\subsubsection*{Load+Reload}
As flush-based attacks do not work with SEV, we need to look at eviction-set based approaches. One particularly efficient method is the AMD-specific Load+Reload attack, as it only needs a single memory access in each stage.
In the original paper~\cite{ASIACCS:LHSPMG20}, the authors only demonstrated the Load+Reload attack in the context of one user space process attacking another.
We were able to reproduce the attack with a malicious hypervisor attacking a regular (non-SEV) VM.
However, when targeting any type of SEV VM (plain, ES, SNP), the observed effect on the cache changes.
Instead of getting an L1 data cache miss and an L2 hit for the evicted address, we observed RAM access times for the \emph{whole memory page} to which the evicted address refers. We were not able to conclusively verify the cause for this behavior.
However, we suspect that this is related to the \enquote{Cache Coherency across encryption Domains} feature~\cite[Sec 15.34.9]{amdProgrammersManualVol2} available on our CPU.
The manual states that without this feature, the HV is required to manually flush a data page of
the VM before accessing it, if it wants to read the latest data. Thus, it is possible that the HV's access to the aliased mapping also internally triggers a cache flush.

\subsubsection*{Prime+Probe}
\label{sec:prime-probe}
As the specialized eviction-set technique of the Load+Reload attack does not work with SEV, we opted for the generic Prime+Probe attack.
To reduce cache noise, we chose to implement the prime and probe steps in the kernel space.
This way, they can be placed immediately before and after the VMRUN instruction.
The eviction set finding itself is implemented in user space, to allow for maximal flexibility.
We found that on our CPU the first 24 bits of the page frame number need to be equal for two pages to be mapped to the same L2 cache set. Next, the user space application passes the virtual addresses of the eviction set(s) to the kernel space component, which will create internal mappings to the used pages. The additional kernel mappings are required as the address space of the user space application will not be mapped during the prime and probe steps performed immediately before and after the VMRUN instruction.

\section{Evaluation}
\label{sec:evaluation}
We evaluated SEV-Step on a Dell PowerEdge R6515 Server with a 3rd generation EPYC 7763 CPU. The attacked VM was protected with SEV-SNP, running Ubuntu 22.10 with an unmodified Linux 5.19.0-26 kernel (starting with 5.19, the mainline Linux kernel supports running as a SEV-SNP guest). The attacker-controlled host is running our modified SEV-Step kernel that is based on AMD's patched Linux 5.14 kernel.
The SEV-Step framework, as well as the code for the evaluation and attacks presented in this paper, is available at
\ifAnon
\url{https://github.com/anonymized}.
\else
\url{https://github.com/sev-step/sev-step}.
\fi

\subsection{Single Step Reliability}
\label{sec:eval-step-reliability}
For the reliability evaluation, we analyzed four different scenarios, based on the ideas described in \autoref{sec:step-window}. The results are shown in \autoref{table:eval-step-reliability}. We define reliability as \enquote{not performing multi-steps} while still performing some single-steps. Starting with an initial guess for the timer value, we iteratively decrease it until any further decrease would result in only performing zero-steps. For all scenarios, we try to single-step a code block consisting of 4000 \texttt{nop} instructions.

\begin{table}[t]
    \caption{Results of single-stepping the same \texttt{nop} slide program while: Resetting the \enquote{accessed} bit before each step (A-Bit), flushing the guest TLB before each step (TLB), doing both (TLB + A-Bit). For the rows with multi-steps, the timer value is the smallest value that did not only produce zero-steps. $\diameter$ M-Step denotes the average amount of instructions executed during a multi-step.}
    \centering
    \begin{tabular}{l c c c c c}
        \toprule
        & Timer & 0-Step & 1-Step  & M-Step & $\diameter$ M-Step\\
        \midrule
        \textsc{Baseline} & 0x31 & 6401 & 1534 & 32 & 37\\
        \textsc{A-Bit} & 0x31 & 6399 & 1548 & 50 &  34\\
        \textsc{TLB} & 0x33 & 1158 & 4000 & 0 & 0 \\
        \textsc{TLB + A-Bit} & 0x33 & 1116 & 4000 & 0 & 0\\
        
        \bottomrule
    \end{tabular}
    \label{table:eval-step-reliability}
\end{table}

In the baseline scenario, we try to achieve single
stepping only using the APIC timer, i.e., without combining it with other (micro)-architectural tweaks. As depicted in the table, this approach fails. Setting the timer value to $0x30$ results in only zero-steps but $0x31$ already gives us $32$ multi-steps.

Next, we analyze the effect of resetting the \enquote{accessed} bit as well as flushing the VM's TLB entries. As explained in \autoref{sec:step-window}, the intention behind these tweaks is to increase and homogenize the timing window leading to a single-step.
Resetting the \enquote{accessed} bit does not have any significant effect. However, flushing the VM's TLB drastically improves the situation, enabling us to execute the targeted program without any multi-steps.
As expected, combining both methods does not yield a significant improvement.

In addition to the slide of \texttt{nop} instructions discussed here, in \autoref{sec:aes-attack}, we single-step the real world Linux kernel AES encryption and decryption code as well as
a more diverse set of instruction microbenchmarks.

\subsection{Event Handling Performance}
To evaluate the performance of the event sending mechanism, we compare handling all events in kernel space with sending them to user space.
As both page faults and single-steps use the same basic mechanism, we restrict our analysis to the code path sending single-step events. We again use the \texttt{nop} slide program introduced in \autoref{sec:eval-step-reliability}.
Without sending events to user space, we require on average $1.007$ ms per single-step event, with a standard deviation of $0.0054$ ms. Sending events to user space requires an average $1.616$ ms per single-step event, with a standard deviation of $0.01548$ ms.
While the user space event handling requires roughly $60\%$ more time, we believe this overhead is acceptable given the substantial improvements in usability and attack development.

\subsection{Cache Attack}
\label{sec:eval-cache-attacks}
We now evaluate SEV-Step's Prime+Probe attack implementation. As discussed in \autoref{sec:prime-probe}, the Load+Reload and Flush+Reload attacks do not work with SEV.
We tested the Prime+Probe attack against both the first level data cache (L1D) and the level two cache (L2). However, as the L1D showed a high amount of noise, we only evaluate the L2 variant here.

For the experiment setup, we assume that the guest physical addresses of both the test program and a given lookup table have already been recovered by the attacker, e.g., by using the page fault side-channel in combination with the \enquote{retired instructions} performance counter~\cite{DBLP:conf/sp/LiWW0TZ22}.
Next, we use page fault tracking to detect when the code is about to be executed, and then start single-stepping to interrupt the code immediately before and after each memory access to the lookup table.
We analyze two variants of a crafted assembly snippet that alternates between accessing offset 64 (byte) and 960 (byte) in a cache line-aligned $16\cdot64$ byte lookup table (similar to the T-tables attacked in \autoref{sec:aes-attack}). 
In the first variant, we placed a \texttt{lfence} instruction between the memory accesses, while for the second variant, the memory accesses are performed back-to-back.
Then we classify the data using a previously determined timing threshold. 
In the first variant, we get a success rate of $0.94$, while for the second variant we only get a success rate of $0.13$.
While the second variant does indeed have higher cache noise, upon closer examination, one of the \enquote{noisy} cache sets is often related to the next upcoming memory access, i.e., despite the fact that we are single-stepping, future memory accesses are already fetched out-of-order and thus leave a cache trace. We discuss these effects in more detail in \autoref{sec:recovering-the-aes-key}, where we demonstrate an end-to-end cache attack against the Linux kernel's AES implementation.
We did not observe any cache trace when zero-stepping an instruction, indicating that the context switch needs to fully complete before any instructions from the VM are issued to the execution pipeline.

\section{Case Studies}
\label{sec:case-studies}
To demonstrate the capabilities of the SEV-Step framework, we performed two case studies. In the first
one, we explore the common workflow of using a SEV VM in combination with an encrypted disk image.
We show how an attacker can use the cache attack and single-stepping features of SEV-Step to recover the
AES volume key of a disk encrypted with LUKS2.
In the second case study, we analyze to which degree SEV-protected VMs are vulnerable to Nemesis-style
attacks~\cite{DBLP:conf/ccs/BulckPS18}. For this, we enrich the single-step events with precise time measurements.
To the best of our knowledge, these kinds of attacks were not explored in the context of SEV before.

\subsection{Cache Attack on Disk Encryption}
\label{sec:aes-attack}
We show an end-to-end, single trace cache attack that is able to steal the volume key of a disk encrypted with cryptsetup+LUKS2, which is a disk encryption system commonly used with Linux. First, we briefly introduce disk encryption, which is a highly relevant workflow
for SEV and confidential VMs in general. Next, we describe how we can force the disk encryption
system to decrypt the disk using a cipher implementation vulnerable to cache attacks.
Finally, we explain the technical details required for gathering the cache traces and how
we recovered the volume key from them.

\subsubsection*{Linux Disk Encryption}

A common approach for deploying SEV VMs is providing the HV with an encrypted disk image and a bootloader. The bootloader is attested through the SEV API and receives the disk password from the user. It then opens the disk image, loads the kernel binary into memory, and transfers control to the kernel. Finally, the kernel unlocks the disk image again and mounts the contained file system. This workflow allows to keep the attested initial code image small, improving performance and reducing the attack surface.
Under Linux, the disk encryption infrastructure~\cite{linuxKernCryptoManual} is split into a user space
and a kernel space part. The kernel contains the disk driver and implementations for several ciphers that can be consumed by user space applications via an API.
An example for this is the popular full disk encryption suite \textit{cryptsetup}.

\textbf{The kernel crypto infrastructure} provides a flexible architecture of basic ciphers and so-called \enquote{templates}.
The former are plain block ciphers (or message digests), the latter implement additional logic on top of existing ciphers. This is commonly used to represent block cipher modes like CBC or XTS.
Part of the kernel crypto API is the \textit{CAPI} specification format, that allows to describe composed ciphers in a structured manner.
For example, \texttt{capi:xts(ecb(aes))-plain64} invocates the XTS driver with AES in ECB mode and a sector number-based IV generator.

Since Linux supports a wide range of architectures, there may be different variants of a cipher, each optimized for a certain architecture version.
Each implementation is assigned two names: The \texttt{cra\_name}, which is equal for all implementations of a given primitive, and the \texttt{cra\_driver\_name}, that uniquely identifies a specific implementation.
The CAPI format supports both names. If a \texttt{cra\_name} is provided, a static scoring system is used to select the best implementation for the current system. If a \texttt{cra\_driver\_name} is specified, the kernel uses that specific implementation if available.

\textbf{The LUKS2~\cite{luks2Spec} format} commonly used with cryptsetup allows specifying the block cipher for the encrypted disk in the CAPI format. As this value is neither encrypted nor authenticated, it can be arbitrarily manipulated, as, e.g., shown in~\cite{cubeOSLUKS2Attack}. The CAPI string is directly passed to the kernel crypto API. We discovered that the Linux kernel shipped with Ubuntu 22.10 contains several symmetric cipher implementations that are highly vulnerable to cache attacks.
In the next section, we show two approaches how a malicious hypervisor can combine these weaknesses, by first tricking the VM into using its secret disk encryption key with a vulnerable algorithm and then extracting the key from the resulting leakage via a cache attack.

\subsubsection*{Forcing Vulnerable Ciphers}
On our test systems, cryptsetup defaults to \texttt{capi:xts(ecb(aes))-plain64}
for LUKS2 encrypted disks. XTS~\cite{ieeeXtsStandard} is a tweaked block cipher mode commonly used for disk encryption. It uses two keys: The first key is used to generate a so-called tweak value by encrypting the current disk sector number. This tweak value, multiplied with a number representing the current offset inside the disk sector, is then XORed to the actual payload data before and after encrypting/decrypting it using the second key.
Using the weaknesses described in the previous section, the malicious HV changes the disk header to \texttt{capi:xts(ecb(cast6))-plain64} before passing the disk to the VM, tricking the VM into using the vulnerable CAST6 implementation.
As the disk content was initially encrypted with AES, this does not yield meaningful plaintext, preventing the disk from being mounted properly. Nevertheless, the decryption routine is still invoked roughly 66k times before the mount operation eventually fails, providing sufficient opportunity to leak the key.

A second, more stealthy approach, is exploiting the ability to force a specific cipher implementation. The attacker replaces the \texttt{capi:xts(ecb(aes))-plain64} specification by \texttt{capi:xts(ecb(aes-generic))-plain64}, such that the AES cipher is instantiated with a leaky T-table based implementation.
While we verified that such substitutions work for \enquote{templates}/composed ciphers, there is one remaining problem when applying it to the XTS implementation in Linux.
Only the cipher instantiation for the payload data encryption/decryption is selected based on the exact value specified in the CAPI string. The cipher instantiation for the tweak generation always uses the priority-based \texttt{cra\_name} to select the best implementation.
As all SEV-enabled systems support AES-NI, this prevents us from leaking the tweak encryption key. Thus, during the key recovery, we cannot recompute the tweak value which is a prerequisite for recovering the second key, used for encrypting the actual payload data.
However, as shown in the next paragraph, a malicious HV can suppress the availability of AES-NI
altogether, forcing the VM to use the vulnerable implementation for both instances of AES.

The VM uses the \texttt{cpuid} instruction to determine whether AES-NI is available. As the HV can intercept this instruction, it can arbitrarily manipulate the reported features.
All versions prior to SEV-SNP cannot detect such manipulations. With SEV-SNP, a new mechanism was added to provide trustworthy \texttt{cpuid} information~\cite{amdSnpAbi}.
During the attestation process, the HV has to commit to a set of \texttt{cpuid} bits, that are additionally verified by the AMD Platform Security Processor (PSP).
Depending on the specific \texttt{cpuid} entry, the PSP enforces different policies.
Some entries are required to match the value on the host, but the AES-NI feature is allowed
to be disabled~\cite[Sec. 2.1.5.3]{amdProcessorProgReferenceFam19hM01hB1}.
Thus, the VM owner has to be aware of the subtle security implications of disabling AES-NI and ensure that their expected attestation value enforces enablement of AES-NI.

As the technical aspects of the cache attack are similar and the AES key recovery is more interesting, we opted for AES in our end-to-end attack. This also leaves the possibility that the VM owner remains unaware of the attack, as the disk mount succeeds.

\subsubsection*{Performing the Cache Attack}

In preparation for the cache attack, we need to solve three challenges: We have to \ding{192} locate the AES code and detect its execution, \ding{193} locate the instructions accessing the AES T-tables, and \ding{194} locate the AES T-tables.

Similar to prior work~\cite{DBLP:conf/eurosys/Morbitzer0HW18,DBLP:conf/sp/LiWW0TZ22,DBLP:conf/uss/LiZWLC21}, we solve \ding{192} through the page fault controlled channel.
As explained before, the decryption of a single XTS-encrypted data block consists of two AES invocations using different keys.
To build the page fault sequence fingerprint, we trace all of the kernel's page accesses
while triggering disk decryption operations. By manual analysis we found
that the page fault sequence in \autoref{table:aes-pf-sequence} uniquely identifies the execution of the relevant AES
functions. While KASLR randomizes the location of the kernel's \texttt{.text} section at each boot, the contents and order of the \texttt{.text} section itself are not randomized.
Furthermore, there are several techniques to break KASLR in the SEV context~\cite{DBLP:conf/sp/MorbitzerPRDS21,DBLP:conf/sp/WilkeWM020}. 
Thus, we encode the page fault sequence relative to the start of the \texttt{.text} section instead
of using absolute addresses, allowing its usage across reboots.
Note that for the final attack, it suffices to track the pages of the sequence one by one, i.e., we no longer need to trigger a page fault on every memory access as we did while generating the fingerprint.

\begin{table}[t]
    \caption{Page fault sequence uniquely identifying the execution of the AES encrypt and decrypt operations performed during the decryption of a single payload data block of the VM's disk. Accesses with the
    \enquote{Marker} role don't correspond to an operation that we want to observe, but are required to accurately
    track the execution flow. The \enquote{PFN Offset} field states the offset of the page containing the function relative to the start of the kernel's \texttt{.text} section (measured in 4096 byte pages).}
    \centering
    \begin{tabular}{l c l}
        \toprule
        Name & PFN Offset & Role\\
        \midrule
        \texttt{xts\_decrypt} & 0x65c & Marker\\
        \texttt{crypto\_cipher\_encrypt\_one} & 0x64b & Marker\\
        \texttt{crypto\_aes\_encrypt} & 0x65f & Tweak generation\\
        \texttt{crypto\_aes\_encrypt} & 0x660 & Tweak generation\\
        \texttt{crypto\_ecb\_decrypt} & 0x65b & Marker \\
        \texttt{crypto\_aes\_decrypt} & 0x660 & Payload decryption\\
        \texttt{crypto\_aes\_decrypt} & 0x661 & Payload decryption\\

        \bottomrule
    \end{tabular}
    \label{table:aes-pf-sequence}
\end{table}

For \ding{193}, we first analyze the assembly code of the AES functions in an offline phase.
This allows us to build a list of all instructions accessing the T-table, each annotated with the number of instructions executed since the start of the function.
Then, during the attack, we single-step the VM's execution once we reach the targeted AES functions. By comparing
the number of executed steps with the information gathered in the offline phase, we know whether we need to perform
the cache attack for the next instruction.

To \ding{194} locate the AES T-tables, we \enquote{sacrifice} the first
memory access instruction of the encrypt/decrypt AES functions: Instead of performing the cache attack, we mark all pages as not present, yielding a list of all pages accessed during the execution of the instruction. We empirically verified that the final page fault before the instruction's retirement corresponds to the page of the T-table.

With the preparation done, we can now single-step the encryption/decryption functions, and perform a L2 Prime+Probe attack on each T-table access. As a T-table has 256 4-byte entries and thus covers 16 cache lines, we need to measure 16 cache sets for each access.

\subsubsection*{Recovering the AES Key}
\label{sec:recovering-the-aes-key}
Given the cache measurements, we now conduct an offline analysis to recover the two AES keys used by the AES-XTS disk encryption. First, we discuss how we overcame the challenge of out-of-order accesses in our cache traces. Afterwards, we describe our key recovery algorithm.

\begin{figure}
    \centering
    \includegraphics[scale=0.6]{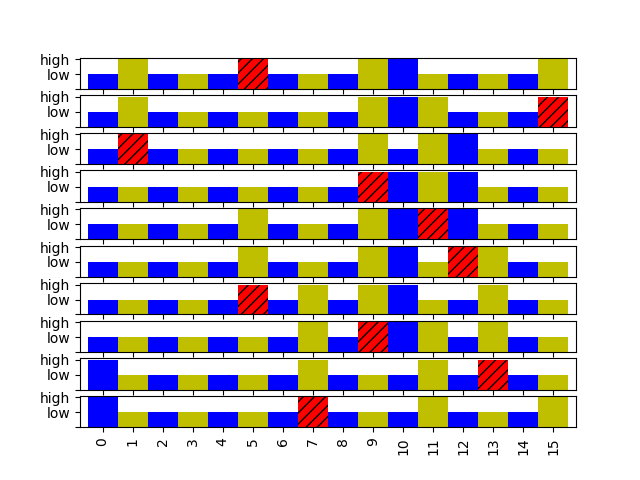}
    \caption{First 10 accesses to the first T-table of \texttt{crypto\_aes\_encrypt}. The X axis shows the cache sets covering the T-table, indexed from 0 to 15. The bars on the Y axis show if the cache set is considered high or low for that access. For each memory access, the actually expected cache set is striped and colored red. Due to out-of-order execution, each expected cache set leaves a trail of high cache sets in the preceding accesses.
   }
    \label{fig:aes-ooo-accesses}
\end{figure}

Although the VM's execution is single-stepped during the cache attack, we still observe a high amount of cache noise, as shown in \autoref{fig:aes-ooo-accesses}. Upon closer examination, most of the cache noise is correlated to future (out-of-order) memory accesses to the same T-table. While those are not actually retired due to the APIC timer interrupt used for single-stepping being processed immediately after the current instruction, their cache traces persist.
This reasoning is supported by our experiment in \autoref{sec:eval-cache-attacks}, where we analyzed a synthetic cache attack victim with and without \texttt{lfences} between the memory accesses and found that the version with fences does not show this behavior. For our attack, a given memory access usually influences up to four preceding accesses to the same lookup table. This matches the round structure of AES, where each T-table is accessed four times with a data dependency between the accesses of different rounds.

As we also have actual cache noise, as well as occasional accesses to the same cache set within four memory accesses, separating the actual access from the noise proved challenging. We opted for a machine learning-based approach with a sequential neural network model consisting of 3 dense layers with 182, 64 and 16 neurons, respectively, as well as two dropout layers to enhance generalization by preventing overfitting.
For the input encoding, we map each memory access to a binary vector, containing the cache traces of the access that we want to classify as well as for the 8 preceding and subsequent accesses. We use 8 instead of 4 accesses in each direction to better improve handling of situations where two close-by memory accesses use the same cache set.
We label each input with a one hot encoding of
the expected memory access. For the first and last 8 accesses, we use zeroes to fill up the missing preceding/subsequent accesses. \autoref{table:aes-ml-classifier} shows the accuracy of the classifier in our experiments. The cache traces for crypto\_aes\_decrypt classification contain a significantly higher amount of noise, leading to a worse classification.
\begin{table}[t]
    \caption{Accuracy of the ML-based classifier for the recorded AES cache traces. We trained a dedicated model
    for each T-table. For crypto\_aes\_encrypt we used approximately 60k training and 11k testing samples.
    For crypto\_aes\_decrypt we used approximately 74k training and 14k testing samples. The varying amount is due to outlier removal.}
    \centering
    \begin{tabular}{l c c c c c}
        \toprule
        Target & Accuracy\\
        \midrule
        crypto\_aes\_encrypt - Lut 0& 0.9050\\ 
        crypto\_aes\_encrypt - Lut 1& 0.9037\\ 
        crypto\_aes\_encrypt - Lut 2& 0.8971\\ 
        crypto\_aes\_encrypt - Lut 3& 0.8610\\ 
        \midrule
        crypto\_aes\_decrypt - Lut 0& 0.6771\\ 
        crypto\_aes\_decrypt - Lut 1& 0.6860\\ 
        crypto\_aes\_decrypt - Lut 2& 0.6919\\ 
        crypto\_aes\_decrypt - Lut 3& 0.2545\\
        \bottomrule
    \end{tabular}
    \label{table:aes-ml-classifier}
\end{table}

For our key recovery, we use XTS decryptions for disk offsets that have known or easily guessable plaintext and that are always accessed during a mount operation. This includes certain magic offsets that are searched for file system headers, and the file system structures themselves. In the first step, we break the key that is used for encrypting the IVs (sector numbers), yielding the tweak. In the second step, we remove the tweak from the ciphertext and then break the key used for decrypting the payload. To break a key, we first guess a number of bits and then check whether that guess is consistent with the T-table measurements, before guessing the next bits. This way, we can discard enough candidates to avoid searching the entire key space. By ordering the candidate cache sets by the probability that is returned by the classifier and discarding measurements with more than 7 candidates, the time needed for finding the correct key can be further reduced.

\subsubsection*{End-to-End Attack}
To test our attack implementation, we created a LUKS2 disk with an \texttt{ext4} filesystem and a random encryption key. We manipulated the header as described to call the vulnerable \texttt{aes-generic} implementation in the kernel, and disabled AES-NI in the VM. When the kernel running inside the SEV VM starts mounting the encrypted disk, we execute steps \ding{192} to \ding{194} to locate the relevant instructions and data structures. We continued with tracing 70 XTS decryptions, from which 34 involved a known plaintext, applied the classifier to the measured cache accesses, and then invoked the key recovery. Our key recovery found the correct key after roughly 13 hours on a 96-core CPU. Note that our attack required a \emph{single mount operation}, making the attack hard to detect and evade.

\subsection{Instruction Latency Attack}
\label{sec:nemesis-attack}
As a second case study, we analyzed whether the interrupt timing-based \emph{Nemesis} attack~\cite{DBLP:conf/ccs/BulckPS18} also applies to SEV.
The core idea of the Nemesis attack is to use the time between single-steps to infer the type of instruction executed, or extract information about its operands. The correlation between the time required for a single-step and the executed instruction stems from the fact that the interrupt used to drive the single-stepping is only processed on instruction boundaries. Thus, the single-step timing correlates to the time required by the executed instruction.
The Nemesis paper analyzed this attack vector for Intel SGX and the Sancus enclave on a TI MSP430 micro controller.
To the best of our knowledge, we are the first to analyze this attack vector on AMD SEV.

\subsubsection*{Measuring Latency}
For measuring the latency of a single-step, we use the \texttt{rdpru} instruction to read the Actual Performance Frequency Clock Count (\texttt{APERF}) MSR, as discussed in \autoref{sec:cache-attacks}. For older Zen processors (prior to Zen 2), the \texttt{APERF} MSR can be read with \texttt{rdmsr} instead of \texttt{rdpru}.
As depicted in \autoref{fig:nemesis-time-measurements}, we obtain a timestamp as close as architecturally possible before and after the VMRUN instruction. The \texttt{sti} instruction in line 14 is required by the virtualization interface and sets \texttt{RFLAGS.IF} to 1, enabling maskable external interrupts. However, it only takes effect after the next instruction has executed, thus our timing measurement cannot be disturbed by interrupts before entering the VM. When leaving the VM, the hardware automatically sets the global interrupt flag (GIF) to 0. This flag disables external interrupts and
thus no such interrupt can trigger between line 15 and 16. As a result, our timestamp code runs in line 19 even before the handler for the APIC timer interrupt that caused the VM exit is executed. The measurement code itself imposes a minimal overhead by storing the timestamps prior to executing \texttt{VMRUN}.

\begin{figure}
    \centering

    \lstdefinelanguage
   [x64]{Assembler}
   {morekeywords={movl, lfence, rdpru, shl, or, mov, vmrun, sti, cli, push},
    numberstyle={\small\ttfamily},
    morecomment=[l]{;}, 
    keywords=[3]{\$32,\$1,
                 eax,ebx,ecx,edx,
                 rax,rdx,rcx,rbx,rsi,rdi,rsp,rbp,r8},
    keywordstyle=[3]\small\bfseries\color{teal!80},
    commentstyle=\small\bfseries\color{violet!70},
    basicstyle={\small\ttfamily}
    }
\begin{minipage}[t]{0.45\textwidth}
\begin{lstlisting}[
    numbers=left,
    language={[x64]Assembler},
    basicstyle=\small
]
; start APIC timer
movl %edx, (%r8)
; timestamp before VMRUN
lfence
movl $1, %ecx
rdpru
shl $32, %rdx
or %rdx, %rax
lfence
; save timestamp to stack
push %rax 
; Prepare VMCB arg
; and enable interrupts
mov %rdi, %rax 
sti

\end{lstlisting}
\end{minipage}
\begin{minipage}[t]{0.45\textwidth}
\begin{lstlisting}[
    firstnumber=16,
    numbers=left,
    language={[x64]Assembler},
    basicstyle=\small,
]
; Enter VM
vmrun %rax
; Execution resumes here
; after VMEXIT
cli

; timestamp after VMRUN
lfence
movl $1, %ecx
rdpru
shl $32, %rdx
or %rdx, %rax
lfence
...
\end{lstlisting}
\end{minipage}

\caption{Assembly code for measuring the time for a single-step event. To reduce system noise to a minimum, we place the time measurement directly inside the kernel space hypervisor code and as close as possible to entering and leaving the VM.}
\label{fig:nemesis-time-measurements}
\end{figure}

\subsubsection*{Differentiating Instructions}
To empirically test the distinguishability of individual x86 instructions based on their latencies, i.e., the difference between the timestamp prior to and directly after \texttt{VMRUN}, we perform experiments in the form of microbenchmarks similar to those of Nemesis~\cite{DBLP:conf/ccs/BulckPS18}. We execute an instruction slide of 1,000 assembly instructions and collect the latencies of each single-step. We repeat this procedure 100 times for a total of 100,000 measurements. Unlike with SGX-Step, we do not need to check the \enquote{accessed} bit in the page table entry to filter for zero-steps, but can directly use performance counters to evaluate the number of zero, single- and multi-steps, as described in \autoref{sec:eval-cache-attacks}. For our analysis, we pick instructions with a range of latencies based on benchmarks done by Abel et al.~\cite{Abel19a}.

\begin{figure}
    \centering
    \begin{subfigure}[t]{0.49\linewidth}\captionsetup{font=footnotesize}
         \centering
         \includegraphics[width=\linewidth]{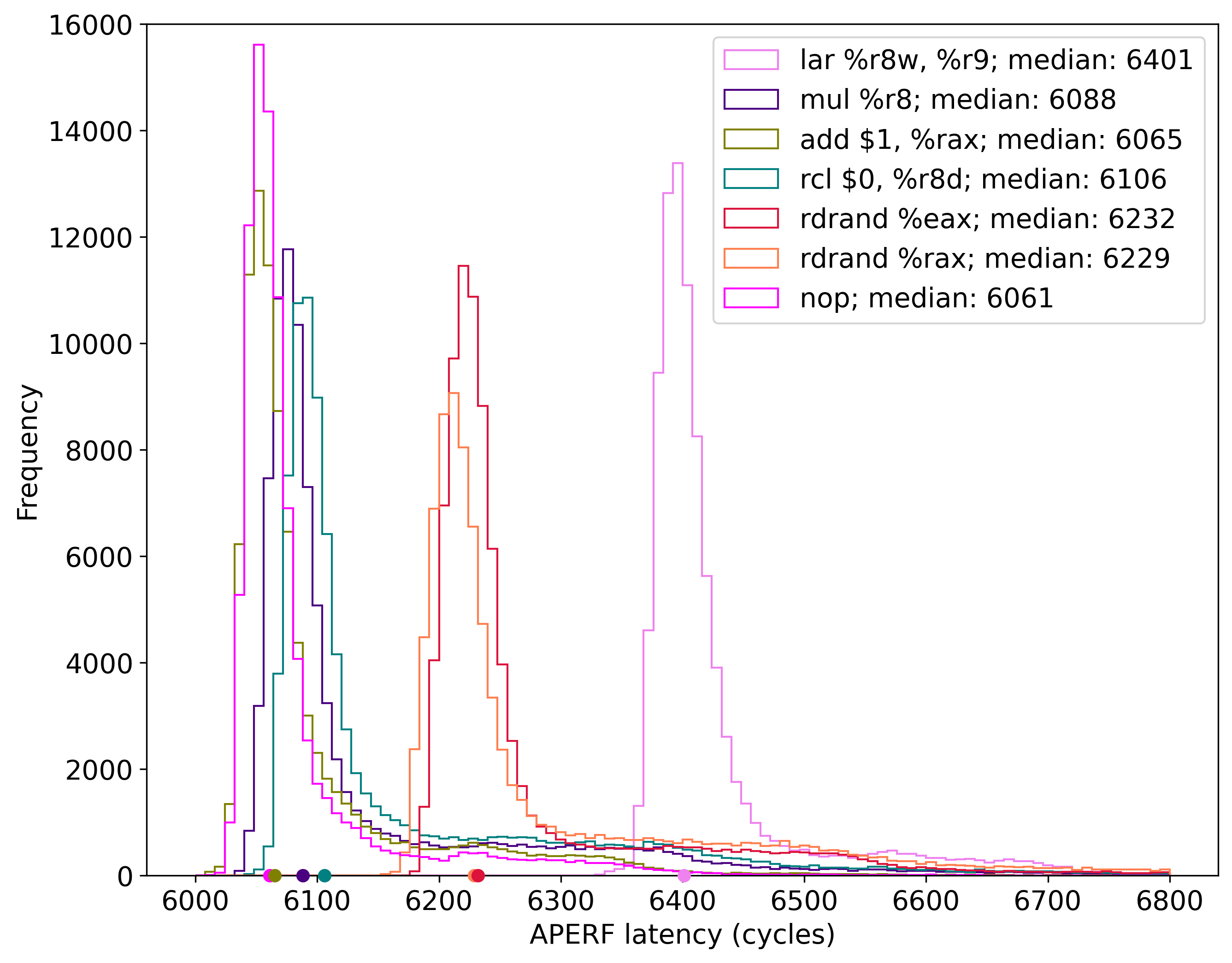}
         \caption{Latency distribution of a selection of x86 instructions.}
         \label{fig:nemesis-instr}
     \end{subfigure}
    \begin{subfigure}[t]{0.49\linewidth}\captionsetup{font=footnotesize}
         \centering
         \includegraphics[width=\linewidth]{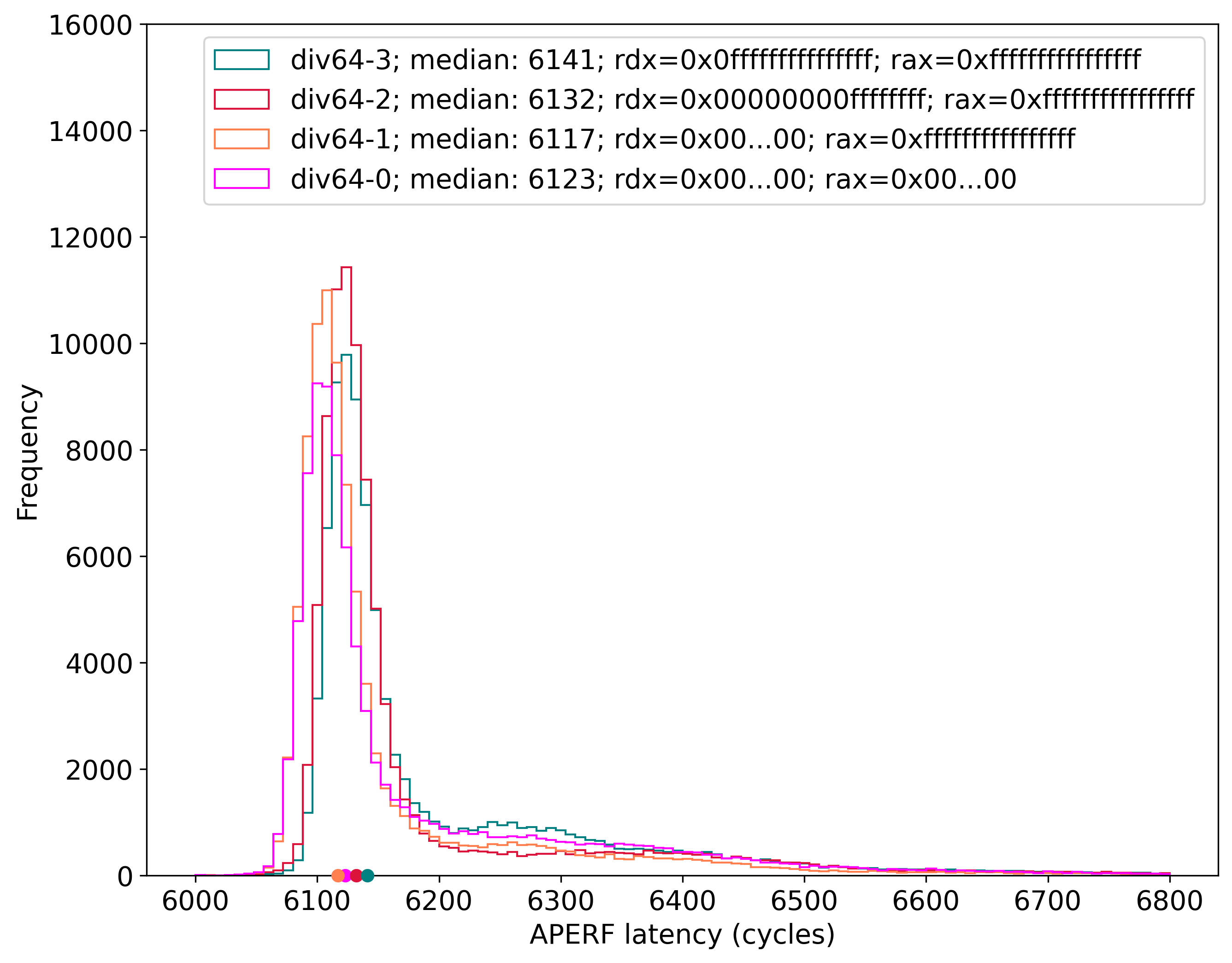}
        \caption{Latency distribution of the x86 \texttt{div} instruction with varying dividend operands and the divisor \texttt{rbx} fixed to \texttt{0xffffffffffffffff}.}
        \label{fig:nemesis-div}
     \end{subfigure}

    \caption{Latency microbenchmarks with 100,000 executions of each instruction.
    }
    \label{fig:nemesis}
\end{figure}

\autoref{fig:nemesis-instr} shows the latency distributions of a selection of x86 instructions. Using SEV-Step, we can distinguish low-latency instructions such as \texttt{add} or \texttt{mul} from high-latency instructions such as \texttt{rdrand} or \texttt{lar}. We also note that, while instructions such as \texttt{nop}, \texttt{add} or \texttt{mul} are harder to discern due to their similar latencies and micro-ops, we can still distinguish the average execution time given sufficient repetitions.

\subsubsection{Differentiating Data Operands}
For determining the distinguishability of data dependent operations, we replicate the experiments of Nemesis regarding unsigned integer division \texttt{div} with the divisor \texttt{rbx} fixed to \texttt{0xffffffffffffffff} while varying the dividend \texttt{rdx:rax}. According to the AMD hardware optimization manual, the \enquote{hardware integer divider unit has a typical latency of 8 cycles plus 1 cycle for every 9 bits of quotient}~\cite[p. 36]{amdSoftwareOptimizationEPYC7003}. This suggests that there should also be a correlation between the significant bits of the dividend and the measured latencies. As can be seen in \autoref{fig:nemesis-div}, while the median latencies for \texttt{div64-1}, \texttt{div64-2} and \texttt{div64-3} increase with the size of the dividend, \texttt{div64-0} does not follow this trend. We also observe that the latency differences between the \texttt{div}s are significantly less prominent than those reported for SGX in the Nemesis paper.

\section{Discussion}
\subsection{Zero/Single-Step Countermeasures}
There are several works that try to protect SGX enclaves against
single-stepping-based attacks~\cite{DBLP:conf/ccs/ChenZRZ17,DBLP:conf/acsac/Lang0MLWL22,DBLP:conf/nordsec/LantzBA22,DBLP:conf/usenix/OleksenkoTKSF18,DBLP:conf/ndss/Shih0KP17}, but none of them
found widespread adoption.
In 2022 Intel in collaboration with researchers~\cite{aexNotify} from
the academic community released the AEX-Notify extensions~\cite{intelAEXWhitepaer}~\cite[p. 199-204]{intelArchInstructionSetExtensionsandFutureFeatures} for SGX that make the enclave interrupt-aware, allowing it to execute custom handler code before resuming at the interrupted instruction. The AEX-Notify paper~\cite{aexNotify} uses this interrupt awareness to execute a code gadget that aims to ensure that the first payload instruction of the enclave will execute fast by ensuring that the instruction as well as its operands are fully cached. This way they aim to prevent reliable zero-/single-stepping.

According to the Intel TDX Module Spec~\cite[Sec 17.3]{intelTDXModuleSpecV1_5Draft}, TDX has been designed with countermeasures for zero-/single-stepping attacks. To prevent single-stepping attacks, a trusted domain (equivalent to SEV VM) can still execute a small randomized amount of instructions if it gets interrupted within approximately 4k cycles after being entered.
To additionally prevent zero-steps via missing page table permissions, the TDX module limits the number of page faults that may occur without forward progress
and thus forces the HV to ensure proper page table configuration before it can resume the trusted domain.

Given the novelty of single-stepping attacks against AMD SEV, we are not aware of any countermeasures. In contrast to the AEX-Notify countermeasure that has to cope with the architectural limitations of SGX, the TDX approach seems more principled. However, in contrast to the TDX design, for SEV there is no trusted layer between the HV and the VM that could e.g. prevent the VM from being entered after a certain amount of faults without forward progress. We leave the design of countermeasures to future work.

\subsection{Preventing Vulnerable Algorithm Selection}
\label{sec:luks-attack-discussion}
As demonstrated in \autoref{sec:aes-attack}, an attacker can exploit the unmodified LUKS2 header in combination with the Linux kernel's expressive CAPI specification language, to trick the VM into decrypting its disk using cryptographic implementations vulnerable to side-channel attacks.
One possible solution is to remove all vulnerable implementations from the Linux kernel, and replace them by constant-time code.
If this is deemed unpractical, the API should flag all vulnerable implementations as such and provide a way to allow its users to explicitly disallow their usage.
Another strategy would be to add a checksum preventing the LUKS2 header manipulation. However, that case  would require to explicitly specify
an implementation for the crypto algorithm. Otherwise, the kernel's priority-based system might still select a vulnerable implementation under certain system configurations.

We used a side-channel leakage analysis tool~\cite{DBLP:conf/ccs/WichelmannSP022} in combination with a custom QEMU plugin to analyze the Linux kernel's crypto primitives for the secret oblivious memory access and constant time properties.
Due to limitations of QEMU, we were not able to analyze AVX-based implementations. We found significant leakages in many other symmetric ciphers, for example
\textit{aes-generic},
\textit{aes-fixed-time},
\textit{blowfish-asm},
\textit{blowfish-generic},
\textit{camellia-asm},
\textit{camellia-generic},
\textit{cast5-generic} and
\textit{cast6-generic}.

We disclosed our findings regarding the LUKS2 header manipulation
and its impact on using LUKS2 in the context of confidential VMs to the cryptsetup/LUKS2 team\footnote{
\ifAnon
    \url{https://gitlab.com/anonymized}
\else
    \url{https://gitlab.com/cryptsetup/cryptsetup/-/issues/809}
\fi
}. As a result, they changed the CAPI parsing part of cryptsetup to disallow the
selection of specific implementations. However, this does not help if all implementations known to the Linux kernel are vulnerable, as it is the case for the blowfish cipher.

\section{Conclusion}

In this paper, we have demonstrated that SEV-SNP VMs can be reliably single-stepped, which greatly increases their vulnerability against a wide range of microarchitectural side-channel attacks. In the hope to ease future research in this direction, we introduced SEV-Step, a reusable framework allowing the development of complex attacks from user space. We have demonstrated the framework's capabilities with two in-depth case studies. 
The cache attack against the Linux disk encryption infrastructure revealed that even with SEV-SNP, the implementation of protected VMs remains brittle due to continuing prevalence of vulnerable code. The clash between the attacker model for which these systems have been designed with their usage in the context of confidential VMs exposes them to powerful software-level attacks in virtualized environments.
Given that not only AMD SEV but also Intel TDX~\cite{intelTDXWhitepaper} and ARM CCA~\cite{arm:cca:2021} employ the confidential VM model, their security under this new threat model should be analyzed with more scrutiny.
Finally, in the second case study, we have demonstrated that SEV is vulnerable to timing-based instruction classification. Like the Nemesis attack on SGX, we were able to confirm that instruction sequences can be reconstructed in SEV. While the timing variation is smaller than in SGX, repeat measurements can reveal even small variations due to data-dependent execution time of instructions such as \texttt{div}.

\bibliographystyle{alpha}
\bibliography{abbrev3,crypto,biblio}

\end{document}